\documentclass[3p,twocolumn,sort&compress,reversenotenum,floatfix]{elsarticle}
\usepackage{graphicx}
\usepackage{subfigure}
\usepackage{dcolumn}
\usepackage{bm}
\usepackage{hyperref}
\usepackage{indentfirst}
\usepackage{braket}
\usepackage{amsmath}
\usepackage{amssymb}
\usepackage{ulem}
\usepackage{color}
\usepackage{makecell}
\usepackage{booktabs}
\usepackage{footnote}
\usepackage[ruled,vlined]{algorithm2e}
\usepackage{tabularx}
\usepackage{lipsum}
\hypersetup{colorlinks=true, citecolor=blue, urlcolor=blue, linkcolor=blue}
\usepackage{chngcntr}
\begin{document}

\title{Ground-state phase diagram of two-component interacting bosons on a two-leg ladder}
\author[1]{Po Chen}
\author[1,2]{Chenrong Liu\corref{cor1}}
\ead{crliu@wzu.edu.cn}
\cortext[cor1]{Corresponding author}
\address[1]{College of Mathematics and Physics, Wenzhou University, Zhejiang 325035, China}
\address[2]{State Key Laboratory of Surface Physics and Department of Physics, Fudan University, Shanghai 200433, China}
\date{\today}
\begin{abstract}
Using the cluster Gutzwiller mean-field method, we numerically study the ground-state phase diagram of the non-hard-core two-component interacting bosons trapped in a two-leg ladder with and without an artificial magnetic field. There are three quantum phases namely Mott insulator (MI), supercounterfluid (SCF), and superfluid (SF) are found in the phase diagram. Interestingly, several loophole SCF phases are observed at a sufficiently small intra- to inter-leg hopping ratio when the magnetic flux is absent. While if the ratio is not so small, the loophole SCF phase would disappear, but it can still be induced by applying a sufficiently large magnetic flux. Additionally, we also find that the presence of the magnetic flux leads to an enlargement of the MI lobe and the conventional SCF lobe. Moreover, the SF-MI phase boundary is quantitatively consistent with the strong-couping expansion at a weak hopping amplitude. 
\end{abstract}

\maketitle

\section{Introduction} 
Ultra-cold atoms in optical lattices offer a versatile and controllable system for investigating fundamental quantum phenomena and simulating complex physical systems. Recent experiment developments of cold atoms in optical lattices allow for studies of the many-body physical properties of Bose-Bose\cite{PhysRevA.77.011603,PhysRevLett.100.210402,PhysRevLett.105.045303,PhysRevA.98.051602}, Fermi-Fermi\cite{science.1122318,science.1122876,PhysRevLett.97.030401,PhysRevLett.97.190407,Nature7179}, and Bose-Fermi \cite{PhysRevA.98.051602,science.1077386,PhysRevLett.92.140405,science.1255380,PhysRevLett.96.180402,PhysRevLett.102.030408} mixtures. The Bose-Bose mixture which is also known as the two-component bosonic mixture can be described theoretically by a two-component Bose-Hubbard model \cite{PhysRevLett.81.3108}. In its ground-state phase diagram, some pairing phases, including the paired superfluid (PSF), the supercounterfluid (SCF), and the molecular superfluid,  can emerge in addition to the superfluid (SF) phase and the Mott insulator (MI) phase \cite{PhysRevLett.92.030403,PhysRevLett.92.050402,PhysRevLett.90.100401,PhysRevA.80.023619,SciPost12,PhysRevLett.125.245301,PhysRevA.108.013309}. The SCF (PSF) is characterized by the formation of interspecies particle-hole (particle-particle) pairs in the presence of repulsive (attractive) interspecies interactions. In this phase, the formed pairings can be regarded as composite bosons, which are in a condensed state, while the single species bosons have no superfluidity \cite{PhysRevA.108.013309}. 

On the other hand, the low-dimensional systems such as the one-dimensional (1D) and quasi-1D systems often attract special interest from researchers because the role the interactions play, which are important for realizing exotic quantum phases, can be enhanced in these systems\cite{RevModPhys.83.1405,PhysRevA.96.053631}. In particular, the two-leg ladder system is more significant since the extra inter-leg coupling can strongly affect the MI-SF phase transitions and the quantum dynamics even in a simple single-component Bose-Hubbard ladder \cite{PhysRevB.63.180508,PhysRevA.96.013620}. Besides, when a magnetic field is introduced, two new quantum phases are observed in the experiment for a single-component Bose-Hubbard ladder, one is the Meissner state and the other is a vortex state\cite{NaturePhysics}. In a small magnetic field and weak interactions, the boson currents circulate along the legs of the ladder resulting in a Meissner state, while for a strong magnetic field, the currents can flow along the rungs and lead to a vortex state. Theoretically, the influence of artificial magnetic fields on the interacting single-component bosonic ladder also has studied by using complex tunnel couplings\cite{PhysRevA.92.023618,Tokuno_2014,Strinati_2018, Orignac_2016, PhysRevA.95.063601,PhysRevB.91.140406,PhysRevResearch.2.043433,PhysRevA.98.063612,PhysRevA.102.053314,PhysRevA.93.053629,PhysRevA.106.063320}.

If a second component is included, the bosonic ladder would then establish a richer ground-state phase diagram due to the additional internal degree of freedom. However, the existing work on two-component bosons in two-leg ladder geometries is much less extensive. Only recently investigations by Jian-Dong Chen \textit{et al}. \cite{PhysRevA.102.043322} have revealed the quantum phases in such a system with a magnetic field, but their study focused on hard-core two-component bosons, where the repulsive intraspecies interaction strength is considered to be infinite.  For this reason,  it is interesting to explore the interplay between the non-hard-core two-component bosons, the magnetic flux, and the effects of the ladder geometry on the properties of the system. Here, we mainly survey the phase diagram of non-hard-core two-component bosons trapped in a two-leg ladder geometry, exploring both cases with and without the presence of an artificial magnetic field.  

This paper is organized as follows. In Sec.\ref{Model}, we describe the two-component Bose-Hubbard model with the artificial magnetic field in a two-leg ladder. In Sec.\ref{Method}, we give some details about the numerical method that we have used, and in Sec.\ref{Results}, we offer the numerical results and some discussions. Finally, in Sec.\ref{Conclusions}, we present a brief summary. 
\section{Model} \label{Model}
\begin{figure}[htbp] 
   \centering
   \includegraphics[width=0.48\textwidth]{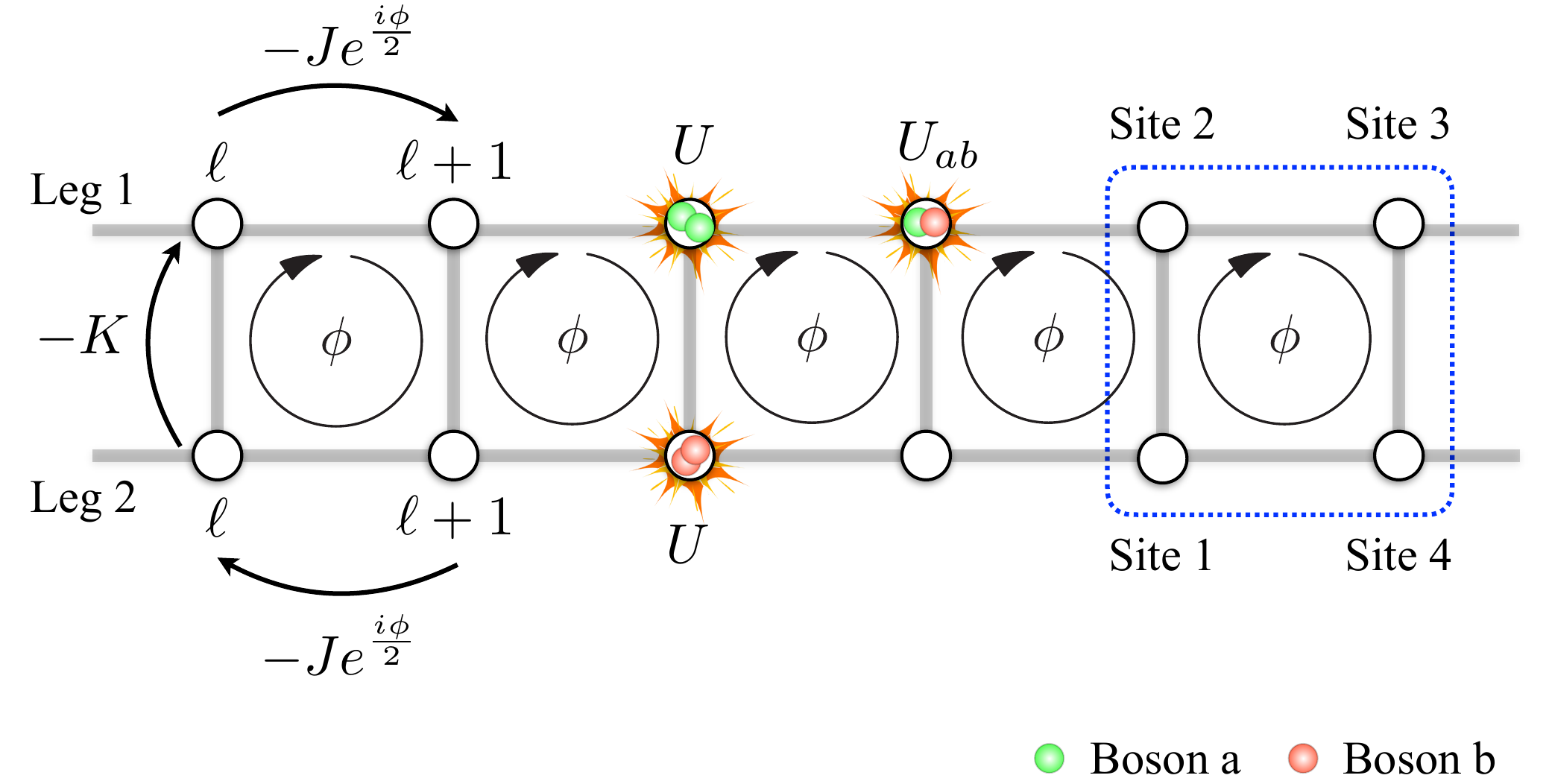} 
   \caption{(Color online) Schematic of the two-component Bose-Hubbard model on a two-leg ladder in the presence of the magnetic field. The dashed blue box indicates the unit cell used for the cluster Gutzwiller mean-field calculations. The sites inside the unit cell are relabeled as 1, 2, 3, and 4 for convenience.  }
   \label{Fig1}
\end{figure}
The Hamiltonian of two-component bosons in a two-leg ladder with interexchange symmetry in the presence of a magnetic field can  be written as
\begin{equation} \label{Ham}
\begin{aligned}
H= & -J \sum_{\ell,\alpha, p} \left[e^{(-1)^p\frac{i\phi}{2}}\alpha_{\ell,p}^{\dagger} \alpha_{\ell+1,p}+\text { H.c. }\right] \\
& -K \sum_{\ell,\alpha}\left[\alpha_{\ell,1}^{\dagger} \alpha_{\ell,2}+\text { H.c. }\right] \\
& +\frac{U}{2}\sum_{\ell,\alpha, p}n_{\ell,p}^\alpha\left( n_{\ell,p}^\alpha -1\right)+U_{ab} \sum_{\ell,p} n_{\ell,p}^an_{\ell,p}^b \\
& - \mu \sum_{\ell,p}(n_{\ell,p}^a+n_{\ell,p}^b),
\end{aligned}
\end{equation}
where $\alpha_{\ell,p}^\dag$ ($\alpha_{\ell,p}$) is boson $\alpha$ ($=a,b$) creation (annihilation) operator at site $\ell$ of leg $p$ ($=1,2$), $ n_{\ell,p}^\alpha$ is the number operator of boson $\alpha$ at site $\ell$ of leg $p$, and $\mu$ is the chemical potential. The  intra- and inter-leg hopping amplitudes are described by $J$ and $K$ respectively, the on-site intra- and inter-species boson repulsive interactions are individually labeled by $U$ and $U_{ab}$. A sketch of this model can be found in Fig.~\ref{Fig1}.  In experiments, the ratios $J/U$ and $K/U$ correspond to the optical lattice laser intensities along each leg and the separation between the legs respectively \cite{PhysRevA.98.063612}. And thus $J/U$ and $K/U$ can be tuned in the experiment. The interspecies interaction $U_{ab}$ can also be varied in experiments via the Feshbach resonances technology\cite{RevModPhys.82.1225}.  

The phase $\phi$ in the intra-leg hopping term is given by,
\begin{equation}
\phi=\oint_{\square} \vec{A}(\vec{r}) \cdot \mathrm{d} \vec{r},
\end{equation}
which is integral in a single plaquette.  Here, $\vec{A}(\vec{r})$ is the vector potential and the magnetic field is then read as $\vec{B}=\nabla \times \vec{A}$. Note that $\phi$ is also the total phase accumulated by the wave function when a particle moves around a plaquette. Due to this reason, $\phi$ is called the gauge flux and it is gauge invariant. The physical properties of Hamiltonian (\ref{Ham}) are therefore gauge invariant and only dependent on the value of $\phi$. In our model, we select a Landau gauge that allows hopping in the rung direction without a gauge field, while hopping along the legs induces a phase. In addition, we assume the two components have the same magnetic field, which means that $\phi_a=\phi_b=\phi$ as illustrated in Fig.~\ref{Fig1}. 

\section{Cluster Gutzwiller mean-field approach} \label{Method}
In this section, we give some details about the cluster Gutzwiller mean-field method (CGMF) \cite{PhysRevA.87.043619} that we used in the present paper. This approach is a powerful tool for studying the interacting bosonic systems. It is an extension of the single-site Gutzwiller approximation by including short-range correlations between different sites within the unit cell which makes the results more reliable. For this reason, the CGMF approach has been successfully applied to the bosonic ladder systems. For example,   R. Sachdeva \textit{et.al.} investigate the ground-state properties of the extended Bose-Hubbard model for two-leg ladder systems in artificial magnetic fields using the CGMF method \cite{PhysRevA.95.063601, PhysRevA.98.063612}, H. Deng \textit{et.al.} present a CGMF study for ground states and time-evolution dynamics in the Bose-Hubbard ladder \cite{PhysRevA.92.023618} and their results are qualitatively consistent with the experimental observation \cite{Nat.Phys.7.61}. 

Note that CGMF is a suitable choice for solving the non-hard-core two-component bosonic mixture,  primarily because of the large degree of freedom of a single site, which is difficult to deal with in other numerical methods. In our calculations, we use a four-site cluster, which is indicated by the blue dashed box in Fig.~\ref{Fig1}, as the unit cell.  We start with the fact that the whole wave function in CGMF can be written as
\begin{equation}
\Psi=|C\rangle|\psi\rangle,
\end{equation}
where $|\psi\rangle$ is the wave function of all sites except for sites within the selected cluster.  $|C\rangle$ is the cluster Gutzwiller trial wave function that can be expressed in terms of the many-site Fock states,
\begin{equation}
|C\rangle=\sum_k c_k |k\rangle,
\end{equation}
where $|k\rangle=|m_{1}^a,m_{1}^b,\cdots, m_{4}^a,m_{4}^b \rangle$ is the Fock state of the chosen cluster (plaquette),  $m_{i}^{\alpha}=0,1,2,3,\cdots, N^c$ is the number of boson $\alpha$ occupied on site $i$, $N^c$ is the maximum $\alpha$-component boson occupy number allowed, $i$ ($=1,2,3,4$) is the index of the site within the plaquette. Notice that we set the two components to have the same value of $N^c$. 

Suppose $|\psi\rangle$ is known, we then project the Hamiltonian (\ref{Ham}) into the cluster many-site Fock basis $\{|k\rangle\}$ and it is
\begin{equation} \label{Mat}
\begin{aligned}
H_{k k^\prime}= & \left\langle\psi\left| H_\psi\right| \psi\right\rangle \delta_{k k^\prime}+\left\langle k\left|H_{C} \right| k^\prime\right\rangle \\
& +\langle\psi|\langle k| H_{\psi C}| k^\prime\rangle| \psi\rangle.
\end{aligned}
\end{equation}
The first term in Eq.(\ref{Mat}) is a constant energy offset and can be discarded. Therefore, the cluster Gutzwiller mean-filed Hamiltonian can be read as 
\begin{equation} \label{HCG}
\begin{aligned}
H_{\rm{CGMF}}= & H_{C}+\langle\psi| H_{\psi C}| \psi\rangle,
\end{aligned}
\end{equation}
where $H_{C}$ is just the exact Hamiltonian (\ref{Ham}) on a $2\times 2$ cluster and $\langle\psi| H_{\psi C}| \psi\rangle$ is the mean-field coupling term,
 \begin{equation}
\begin{aligned}
\langle\psi| H_{\psi C}| \psi\rangle=&-J \sum_{\alpha=a,b}\left[e^{\frac{-i \phi}{2}} (\alpha_{1}^{\dagger} \chi_\alpha+\alpha_{4} \gamma_\alpha^* )+ \right. \\ 
& \left. e^{\frac{i \phi}{2}} (\alpha_{2}^{\dagger} \beta_\alpha+\alpha_{3} \eta_\alpha^* ) +\text { H.c. }\right]
\end{aligned}
\end{equation}
where $\alpha_s$ ($s=1,2,3,4$) is the $\alpha$-component boson annihilation operator on the site $s$. Here, we introduce the plaquette superfluid order parameter as 
 \begin{equation}
\Omega_{\alpha}=(\chi_\alpha, \gamma_\alpha,  \beta_\alpha, \eta_\alpha).
\end{equation}
In CGMF, $\Omega_{\alpha}$ can be solved by using a self-consistent iteration procedure. The main steps are:
\begin{itemize}
\item Step 1: Initializing a random value of  $\Omega_{\alpha}$;
\item Step 2: Substituting $\Omega_{\alpha}$ into the Hamiltonian matrix (\ref{HCG}). Then, diagonalizing the matrix to obtain the ground state $|C\rangle$ and the corresponding ground-state energy $E$;
\item Step 3: Replace $\Omega_{\alpha}$ with the following new values,
 \begin{equation}
 \begin{aligned}
\chi_\alpha &=\langle C | \alpha_4 |C \rangle , \gamma_\alpha =\langle C| \alpha_1 |C\rangle \\
\beta_\alpha &=\langle C| \alpha_3 |C\rangle, \eta_\alpha =\langle C| \alpha_2 |C\rangle
\end{aligned}
\end{equation}
 where $\alpha=a,b$;
\item Step 4: Repeat steps 2 and 3 until convergence is achieved. 
\end{itemize}

Although the above steps are simple, some tricks still need to be dealt with carefully. Firstly, the dimension of the Hamiltonian matrix (\ref{Mat}) is $D=(1+N^c)^{8}$. This means $D$ growth polynomially with $N^c$.  For example, if we set $N^c=2$, then $D=6561$, while if $N^c=4$, then $D=390625$. Due to this large matrix dimension, the full matrix store scheme is impossible to apply. We should thus use a sparse matrix format in which only the non-zero elements are stored.  Secondly, the numerical results would be more reliable when $N^c$ is sufficiently large. To determine the suitable choice of $N^c$, we have tested the measurements for several values of $N^c$ in \ref{A} and find that $N^c=3$ is good enough when $\mu/U$ is lower than 1.50.  For this reason, in most of our calculations, $N^c$ is set to be 3, meaning that $D=65536$. When $\mu/U$ comes from 1.0 to 1.5, $N^c$ is set to be 4. This alternately setting of $N^c$ can improve the computational efficiency as well as the validity of the numerical results. Lastly, the matrix dimension is too large to apply the usually full matrix diagonalization algorithm, and we thus need to perform the high-precision Arnoldi iteration, which has been integrated into the arpack library \cite{arpack}. This algorithm is a powerful technique for obtaining the lowest few eigenvalues and eigenstates of the large sparse matrix.  

In addition, the convergence condition is set as $\Delta<10^{-6}\sim10^{-8}$, where $\Delta$ is the energy and order parameter difference between two continuous self-consistent iterations,
 \begin{equation}
 \begin{aligned}
\Delta=&\left|E_{d} / U-E_{d-1} / U\right|+\sum_{\alpha=a,b; i}| |\Omega_\alpha^d(i)|-|\Omega_\alpha^{d-1}(i)||. 
\end{aligned}
\end{equation}
Here, $d$ indicates the iteration index within a self-consistent loop, $\Omega(i)$ represents the $i$-th ($i=1,2,3,4$) site superfluid order parameter. When a self-consistent loop is finished, the cluster Gutzwiller wave function $|C\rangle$ and the plaquette order parameters $\Omega_\alpha$ would be both determined. Due to this reason, the initial knowledge of $|\psi\rangle$ is not required.

\section{Results} \label{Results} 
Below we present the ground-state properties with and without an artificial magnetic field under a fixed value of the interspecies interaction $U_{ab}/U=0.5$ (in units of $U$).  Remind that this value has no special meaning, it only affects the chemical potential width of the SCF lobes at $J=0$ \cite{PhysRevA.108.013309}.  

\subsection{Without the magnetic field case: $\phi=0$} \label{RA}

\begin{figure*}[htbp] 
 
\twocolumn[
{ \centering
   \includegraphics[width=0.9\textwidth]{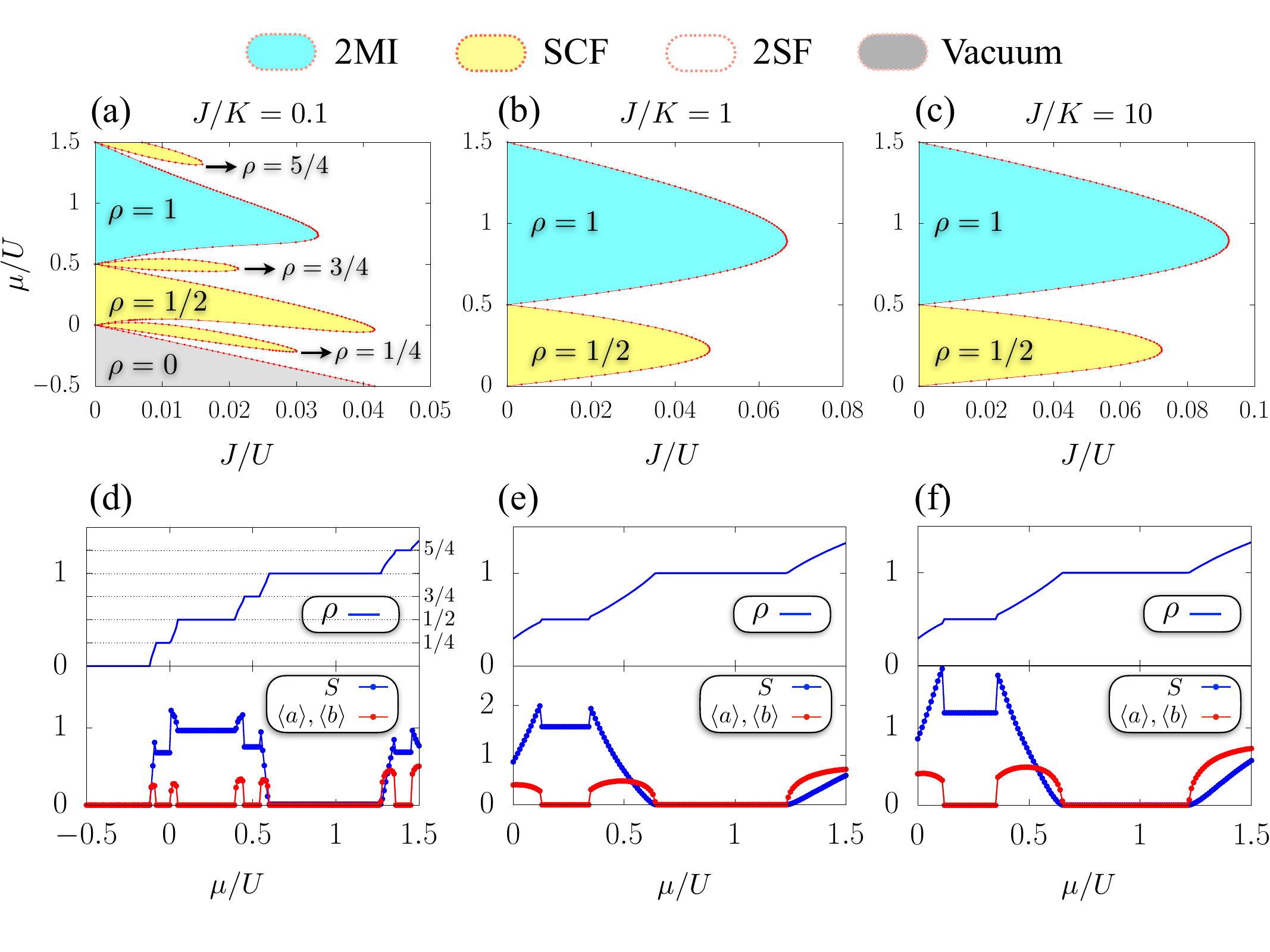}  
   \caption{(Color online) CGMF ground-state phase diagrams and measurements for the two-leg ladder two-component Bose-Hubbard model in the absence of the magnetic field. (a)-(c) $J/U$-$\mu/U$ phase diagram for three different values of $J/K$, and their corresponding measurements as a function of $\mu/U$ are presented in (d)-(f), where $J/U$ is fixed at $0.01$, $0.04$, and $0.06$ respectively. Here, $\rho$ is the single component averaged particle numbers per site, $\langle a \rangle$ and $\langle b \rangle$ are the superfluid order parameters for boson $a$ and $b$ respectively, $S$ is the interspecies entanglement entropy which can be used to identify the SCF phase.  At $\phi=0$,  $\langle n^a \rangle=\langle n^b \rangle =\rho$ and $\chi_\alpha=\gamma_\alpha=\beta_{\alpha}=\eta_{\alpha}=\langle \alpha \rangle$ ($\alpha=a,b$). In (a)-(c), different colors are applied to label different phases, while the 2SF phase remains uncolored. } \label{Fig2} 
   }
   \vspace{1cm}
   ]
\end{figure*}

For a non-magnetic field situation, the model is reduced to the ordinary two-component Bose-Hubbard model in a two-leg ladder. In this simple scenario, the system has several symmetries which are preserved. For instance, the interexchange symmetry, the up-down symmetry, and the left-right symmetry of the ladder geometry.  As a result, the plaquette superfluid order parameters are real numbers and satisfy the following relations,
 \begin{equation}
\chi_\alpha=\gamma_\alpha=\beta_{\alpha}=\eta_{\alpha}=\langle \alpha \rangle,
 \end{equation}
 where $\alpha=a,b$ and $\langle a \rangle=\langle b \rangle$. The $\alpha$-component superfluid order parameters thus can be simply written as $\langle \alpha \rangle$. We use this denotation to show the corresponding order parameters in Fig.~\ref{Fig1} which we will discuss later.

We remind here that although the considered system is in one (quasi-one) dimension, a beyond mean-filed study has shown that the superfluid phases such as the 2SF and SCF can still exist\cite{PhysRevA.80.023619}. Before discussing the phase diagram, let us talk in some detail about how we identify the different phases.  To determine the 2SF-2MI phase boundary, people usually use the value of $\langle \alpha \rangle$, i.e. in the 2SF phase, $\langle a \rangle$ and $\langle b \rangle$ are both non-zero, while they are both equal to zero in the 2MI phase. But for the SCF phase,  both $\langle a \rangle$ and $\langle b \rangle$ also vanish. Hence, $\langle a \rangle$ and $\langle b \rangle$ are insufficient to distinguish the SCF phase from the 2MI phase, we therefore need another parameter. In the single-site Gutzwiller mean-field method, this extra order parameter is simply $\langle ab^\dag \rangle$, e.g. in a SCF state, $\langle a \rangle=\langle b \rangle=0$ and $\langle ab^\dag \rangle \neq 0$. However, $\langle ab^\dag \rangle $ can not be used as the SCF order parameter in the CGMF calculations. This is because $\langle a \rangle=\langle b \rangle=0$ implies that the second term of the Hamiltonian (\ref{HCG}) is absent, leaving only $H_C$ where the total particle number of the single component is conserved. Thus, the onsite order parameter $\langle ab^\dag \rangle $ should always equal zero in the cluster ground state of the SCF phase.  

Fortunately, according to our previous work, the interspecies entanglement entropy $S$ can effectively distinguish between an SCF state and a 2MI state \cite{PhysRevA.108.013309}. Here, $S$ refers to the quantum entanglement entropy between the two species and is defined as 
 \begin{equation}
S=-\sum_k\left|q_k\right|^2 \ln \left|q_k\right|^2
 \end{equation}
where $q_k$ is the $k$-th singular value of the singular value decomposition of the reshaped ground-state matrix. Here, $S$ is calculated using the species partition instead of the space-like or lattice partitions (See details in \ref{B}). Because in a SCF state, the two components are entangled with each other and we expect that $S$ is non-zero in this phase. But for a $\rho=m$ 2MI lobe, the cluster ground state is 
 \begin{equation}
|\Psi\rangle=|m_1^a, m_1^b, \cdots, m_4^a, m_4^b\rangle=|m, m, \cdots, m, m\rangle,
 \end{equation}
which makes $S$ to be zero. Based on these discussions, a summarization of the behavior of the physical quantities in different phases has been made in Table \ref{table1}. The resulting phase diagrams are shown in Figs.~\ref{Fig2}(a)-(c), and the corresponding measurements are presented in Figs.~\ref{Fig2}(d)-(f). 

\begin{table}[tb]
\caption{\label{table1}%
Identification of 2SF, SCF, and 2MI phases in terms of the values of superfluid order parameters $\langle \alpha \rangle$ ($\alpha=a,b$), interspecies entanglement entropy $S$, and single-component averaged particle number per site $\rho$. Here, Int. labels the integer.
}
\vspace{0.7mm}
\small
\begin{tabularx}{\linewidth}{XXXX}
\toprule
Phases &
$\langle a \rangle$, $\langle b \rangle$ &
$S$ &
$\rho$\\
\midrule
\midrule
2MI & =0 & =0 & Int.\\
SCF & =0 & $\neq 0$  & Non-int.\\
2SF & $\neq 0$ & $\neq 0$ & Non-int.\\
\bottomrule
\end{tabularx}

\end{table}

Interestingly, when the intra- to inter-leg hopping ratio $J/K$ is sufficiently small, i.e. $J/K=0.1$, several loophole SCF phases appear around the conventional SCF lobe in the $J/U-\mu/U$ phase diagram as depicted in Fig.~\ref{Fig2}(a). The appearance of these loophole SCF phases is a direct result of $U\gg J$, $U_{ab}\gg J$, and $K\gg J$, which means that,
\begin{itemize}
\item when $U\gg J$, the tunneling along the leg direction is suppressed, and thus the superfluid order parameter vanishes;
\item since $K\gg J$, the particles may still freely move along the rung.
\item because $U_{ab}\gg J$, the $ab$ particle-hole parings can be formed and a hole can also move along the rung together with its paired particle. 
\end{itemize}
Based on these three points, we can immediately deduce that in the simplest SCF ground state at $J/K=0.1$, there is only a single particle that can freely move along the rung. This particle is from either boson $a$ or boson $b$ and is paired with a hole from another component. This hole is moved together with its partner particle. Due to $U\gg J$, this particle-hole paring is localized in a rung.  Consequently, the averaged single-component particle number per site in this state should be $\langle n^a \rangle= \langle n^b \rangle =\rho=1/4$.  This matches the $\rho=1/4$ loophole SCF phase. Due to the particle-hole symmetry being preserved, it is expected that in addition to the $\rho=1/4$ loophole SCF phase, there would also exist a $\rho=3/4$ loophole SCF phase. Indeed, the calculation of $\rho$ as a function of $\mu/U$ for a fixed value of $J/U$ in Fig.~\ref{Fig2}(d) has confirmed this.    

The loophole shape of these special SCF phases can be explained as follows. When $J\rightarrow 0$, $K=J/0.1$ is approached to zero and thus all the sites are decoupled. This means that the phase diagram at $J=0$ is similar to that of the single-site Gutzwiller method, with only two phases: the conventional SCF and the 2MI lobes. As a result, the loophole SCF phase can not appear at $J=0$, which means the region of this phase is surrounded by the regions of the other phases and its phase boundary would be a closed curve.  It is consistent with the numerical results presented in Fig.~\ref{Fig2}(a). Furthermore, Fig.~\ref{Fig2}(d) demonstrates that the interspecies entanglement entropy $S$ remains non-zero in the loophole SCF phases. This is attributed to the presence of interspecies particle-hole pairings in the loophole SCF phase, which highlights a significant distinction from the 2MI phase, despite they both have a zero atomic superfluid order parameter.  

Besides, it should be mentioned that several loophole shape Mott-insulator (LMI) phase regions can also be found in the phase diagram of the single-component bosonic ladder \cite{PhysRevA.92.023618} and superlattice \cite{PhysRevA.70.061603,PhysRevA.72.031602,PhysRevLett.110.260405,PhysRevA.72.013614,PhysRevA.94.023634} systems.  In the LMI phase, the averaged particle number is a non-integer rather than an integer, which is distinguished from the usual MI phase. Because there are only single-component bosons, parings can not be formed in the LMI phase. This is essentially different from our loophole SCF phases. 

On the other hand, if the intra- to inter-leg hopping ratio $J/K$ is large, then the two-leg ladder can be regarded as two decoupled chains and the phase diagram is therefore similar to that of a two-component Bose-Hubbard chain where there are only conventional SCF lobes, 2SF phase, and 2MI lobes. To show this,  the phase diagrams for $J/K=1$ and $J/K=10$ are presented in Fig.~\ref{Fig2}(b) and Fig.~\ref{Fig2}(c) respectively. One can see from the figures that there are only known $\rho=0.5$ lobe and $\rho=1$ 2MI lobe. To recognize these two lobes, the related measurements have been shown in Figs.~\ref{Fig2}(e) and (f). Moreover, as $J/K$ is increased from 1 to 10, we find that the areas of the SCF lobe and the 2MI lobe have a tiny expansion. This is because the weakening of intra-leg tunneling occurs when the assistance of inter-leg tunneling is suppressed.

\subsection{With the magnetic field case: $\phi\neq 0$}
\subsubsection{The single-particle energy band structure}
\begin{figure}[t] 
   \centering
   \includegraphics[width=0.4\textwidth]{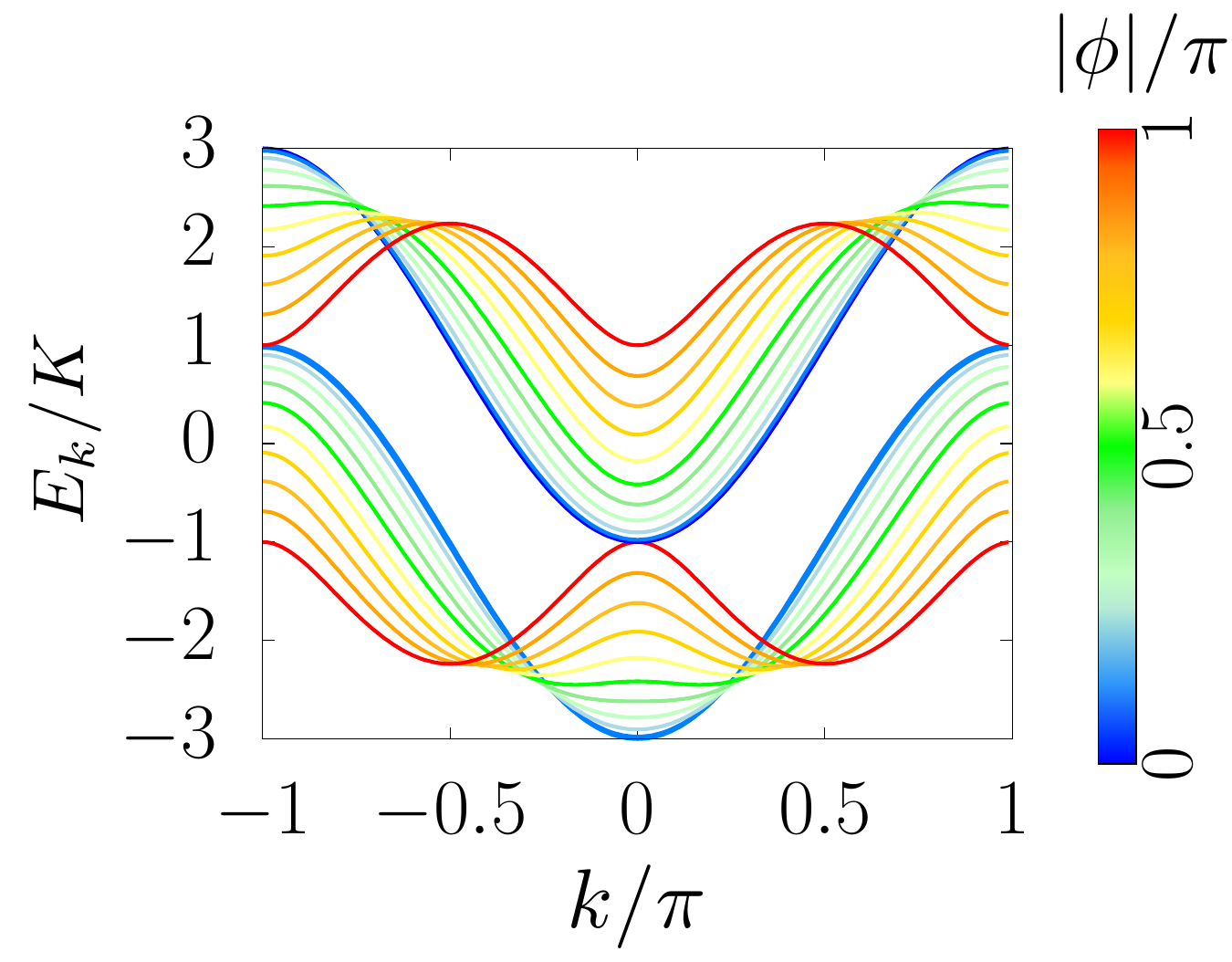} 
   \caption{(Color online) The single component single-particle energy band structure of the two-leg ladder for a varied absolute magnetic flux under a fixed intra- to inter-leg hopping ratio $J/K=1$. Note that the two species have the same single-particle energy band structure. }
\label{Fig3}
\end{figure}
So far, we have investigated the ground-state properties of the two-component bosonic ladder in the absence of the magnetic field.  In this section, we are interested in studying the effect of the magnetic field on the phase diagram. To facilitate further discussion, we produce the single-particle energy band structure for various magnetic flux values in this part. Although our system consists of two-component bosons, the single-particle energy band structure is identical for both species because bosons $a$ and $b$ are equivalent. To see this, we set $U=U_{ab}=0$ and performing the following Fourier transformation,
\begin{equation} 
\begin{aligned}
\alpha_{\ell,p}^\dag=\frac{1}{\sqrt{L}} \sum_k \alpha_{k,p}^\dag e^{i k \ell}
\end{aligned}
\end{equation}
where $\ell$ is the site index in a single leg, $p=1,2$ is the leg index, $L$ is the total number of lattice sites in a single leg, and $k$ is the reciprocal lattice vector.  Here, $\left[\alpha_{k,p}, \alpha_{k^\prime,p^\prime}^\dagger\right]=\delta_{k, k^{\prime}}\delta_{p, p^{\prime}}$. The Hamiltonian (\ref{Ham}) is then takes the form,
\begin{equation} 
\begin{aligned}
H_k &=\sum_{k} \Psi_k^\dag h_k  \Psi_k \\
h_k &= \left( \begin{matrix} A & B & 0 & 0 \\ B & C & 0 & 0 \\ 0 & 0 & A & B \\ 0 & 0 & B & C\end{matrix} \right)
\end{aligned}
\end{equation}
and,
\begin{equation} 
\left\{
\begin{aligned}
 \Psi_k^\dag&=(a_{k,1}^\dag a_{k,2}^\dag, b_{k,1}^\dag, b_{k,2}^\dag), \\ A &=-2J\cos(\phi/2+k),\\ B & =-K, \\ C &=-2J\cos(\phi/2-k).
\end{aligned}
\right.
\end{equation}
When diagonalizing $h_k$ is achieved,  the single-particle energy can be found as,
\begin{equation} \label{Ek}
E_k^{\pm}=\frac{1}{2}\left(A+C\pm\sqrt{A^2+4B^2-2AC+C^2}\right).
\end{equation}
The above expression reveals a two-band structure and $2\pi$ periodicity in the momentum space which indicates the two-leg ladder lattice geometry.  In Fig.~\ref{Fig3}, we plot the single-particle energy band structure for several values of $\phi$ under a fixed value of $J/K=1.0$. 

One can see that for small values of $|\phi|$, i.e. $|\phi|<\phi_c$, the minimum of the lower band is situated at $k=0$, but when $|\phi|>\phi_c$, it evolves into two degenerate and symmetric minima located at $k=\pm k_0$. As a result, the system exhibits two distinct ground states for low and large values of $|\phi|$,  which have been experimentally realized for single-component weakly interacting bosons and named the Meissner and vortex phases respectively \cite{NaturePhysics}.  Here, $\phi_c$ and $k_0$ can be read as \cite{PhysRevA.91.013629},
\begin{equation}  \label{crt}
\left\{
\begin{aligned}
\phi_{\mathrm{c}} &=2 \arccos \left(-\frac{K}{4 J} \pm \sqrt{\left(\frac{K}{4 J}\right)^2+1}\right) \\
k_0 &=\arcsin \left(\sqrt{\sin ^2\left(\frac{\phi}{2}\right)-\frac{K^2}{4 J^2 \tan ^2\left(\frac{\phi}{2}\right)}}\right).
\end{aligned}
\right.
\end{equation}
These two quantities are used to determine the lowest value of the single-particle energy band which is an important parameter in the strong-coupling expansion presented below. Although $J/K$ is fixed at 1.0 in Fig.~\ref{Fig3}, we have checked and found similar results for other values of $J/K$ as well. The only change is that the single bandwidth is increased as $J/K$ increases. 

\begin{figure*}[tb] 
   \centering
   \includegraphics[width=0.9\textwidth]{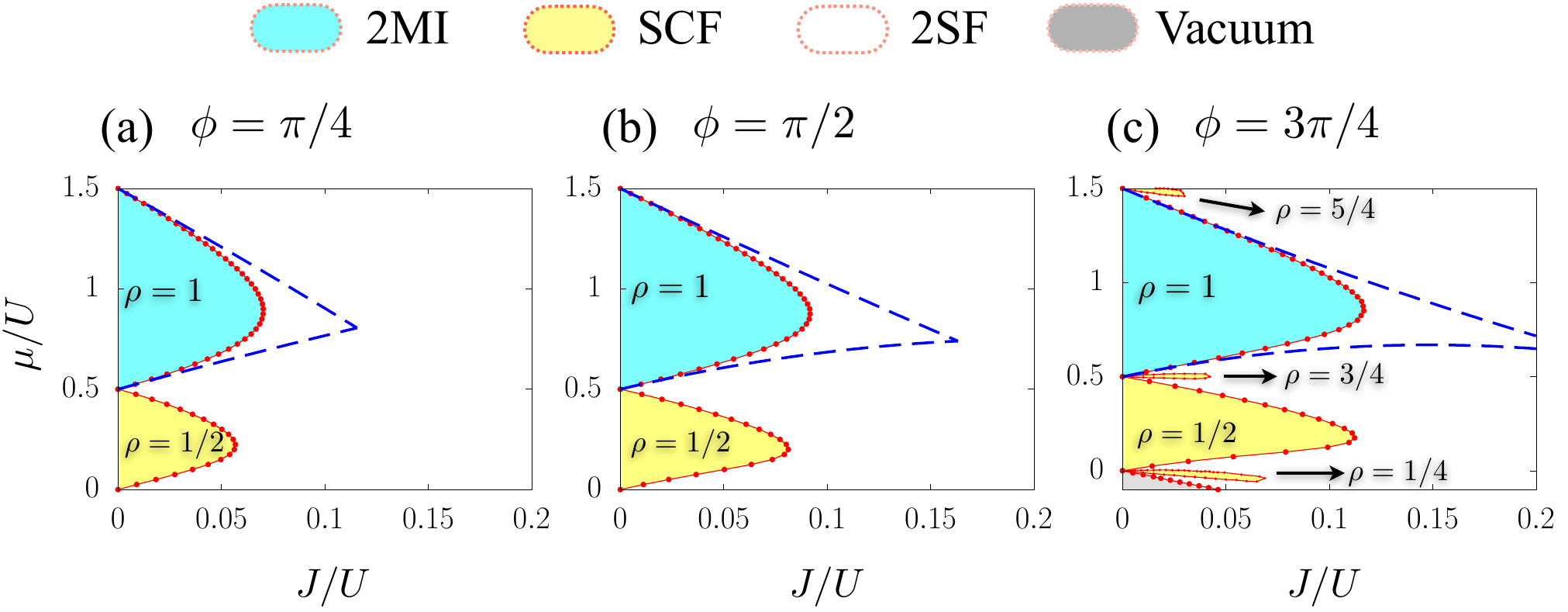} 
   \caption{(Color online) CGMF ground-state phase diagrams for the two-leg ladder two-component Bose-Hubbard model with three different values of magnetic flux at a fixed ratio $J/K=1$. Here, different colors are applied to label different phases, while the 2SF phase remains uncolored. The blue dashed lines are the 2MI-2SF phase boundaries obtained using the strong-coupling expansion technique.}
\label{Fig4}
\end{figure*}

\subsubsection{Strong-coupling expansion}
When the interaction is considered, the system can be in three quantum phases which are SCF, 2MI, and 2SF as we have discussed. Among this, the 2MI-2SF phase boundary can be approximately and analytically determined by performing the strong-coupling expansion technique which is a many-body version of the Rayleigh-Schr\"odinger perturbation theory in the hopping term. For the single-component Bose-Hubbard model, the strong-coupling expansion has been applied successfully in low dimensions and has been proven to be in good agreement with the numerical method \cite{PhysRevB.61.12474,PhysRevB.58.R14741}. However, for the two-component Bose-Hubbard model, this method is unable to locate the SCF-2SF phase boundary \cite{PhysRevA.82.033630}. Therefore, we can only give the phase boundary of the transition from the 2MI to the 2SF phase in this part.  

In the spirit of the strong-coupling expansion, one needs to know the unperturbed perfect 2MI states, and the defect state by adding an extra particle or hole into the perfect 2MI states, and then compare the energy of the 2MI state with that of the defect state. The details of this method can be found in Refs.\cite{PhysRevA.91.013629,PhysRevA.82.033630} and we simply give the main formulas here. For using the perturbation theory, let us rewrite the model Hamiltonian (\ref{Ham}) in the generalized form,
\begin{equation}
H =H_T+H_{int},
 \end{equation}
where $H_T$ and $H_{int}$ are,
\begin{equation} 
\begin{aligned}
H_T=&\sum_{i, j}T_{i, j}  \left(a_i^{\dagger} a_j+b_i^{\dagger} b_j\right) \\ 
H_{int}=&\frac{U}{2} \sum_i \left(n_i^a\left(n_i^a-1\right)+n_i^b\left(n_i^b-1\right)\right)\\ &+\frac{U_{ab}}{2} \sum_i n_i^an_i^b -\mu \sum_i \left(n_i^a+n_i^b\right).
\end{aligned}
 \end{equation}
Here, $T_{i,j}$ is a complex conjugate matrix that is related to the lattice structure, and $H_T$ is treated as the perturbation Hamiltonian. The perturbative energies of the 2MI phase $E_{\rm{2MI}}^{(2)}$, the additional single-particle state $E_{\rm{2MI},+}^{(2)}$, and the additional single-hole state $E_{\rm{2MI},-}^{(2)}$ up to second order are found to be
\begin{equation}
\begin{aligned}
E_{\rm{2MI}}^{(2)} = &N_{\mathrm{s}}[U\rho\left(\rho-1\right)+U_{ab}\rho^2-2\mu \rho  \\ & -\frac{2(2 J^2+K^2)}{U} \rho\left(\rho+1\right)] ,\\
E_{\rm{2MI},+}^{(2)} =&E_{\rm{2MI}}^{(2)}+U\rho+U_{ab}\rho-\mu \\ &+\lambda_T(\rho+1) -\frac{\lambda_T^2}{U}\rho(\rho+1)\\ &+\frac{(2 J^2+K^2)}{2U} \rho\left(5\rho+4\right), \\
E_{\rm{2MI},-}^{(2)}=&E_{\rm{2MI}}^{(2)}-U(\rho-1)-U_{ab}\rho+\mu \\ &+\lambda_T\rho -\frac{\lambda_T^2}{U}\rho(\rho+1) \\ &+\frac{(2 J^2+K^2)}{2U} (\rho+1)\left(5\rho+1\right),
\end{aligned}
\end{equation}
where $N_s$ is the total number of sites, $\rho$ is the single-component particle number per site in a 2MI phase ($\langle n^a \rangle=\langle n^b \rangle=\rho$), $\lambda_T$ is the lowest eigenvalue of the hopping matrix $T_{i,j}$ which is just the lowest value of the single-particle energy band and can be obtained through the Eq.(\ref{Ek}) and Eq.(\ref{crt}). Note that since there are two components, the additional particle (hole) can either be from boson $a$ or from boson $b$ and this does not change the energies of the two defect states because of the interexchange symmetry. Solving the equations $E_{\rm{2MI},+}^{(2)}=E_{\rm{2MI}}^{(2)}$ and $E_{\rm{2MI},-}^{(2)}=E_{\rm{2MI}}^{(2)}$ for $\mu$ separately, the 2MI-2SF phase boundary is then given by,
\begin{equation}\label{st2MI}
\begin{aligned}
\mu^+ =&U\rho+U_{ab}\rho +\lambda_T(\rho+1) -\frac{\lambda_T^2}{U}\rho(\rho+1)\\ &+\frac{(2 J^2+K^2)}{2U} \rho\left(5\rho+4\right), \\
\mu^-=&U(\rho-1)+U_{ab}\rho-\lambda_T\rho +\frac{\lambda_T^2}{U}\rho(\rho+1) \\ &-\frac{(2 J^2+K^2)}{2U} (\rho+1)\left(5\rho+1\right).
\end{aligned}
\end{equation}
When $U_{ab}=0$, the above phase boundary recovers the known strong-coupling expansion result for the single-component Bose-Hubbard ladder with magnetic flux 
\cite{PhysRevA.91.013629}. Remind that here the magnetic effect has been indirectly considered in the eigenvalue $\lambda_T$. This phase boundary can be used to compare with our CGMF calculations, which is the subject of the next section. 

\subsubsection{CGMF ground-state phase diagram for different magnetic flux}
Now let us examine how the magnetic flux affects the ground-state phase diagram of the two-component Bosonic ladder by using the CGMF. For simplicity, the ratio $J/K$ is fixed at 1 in this section and the $J/U-\mu/U$ phase diagram is plotted for $\phi=\pi/4$, $\phi=\pi/2$, and $\phi=3\pi/4$ in Figs.\ref{Fig4}(a),(b), and (c) respectively. As it is shown, a main feature of the phase diagram is that an increase in the magnetic flux results in the expansion of the regions of the 2MI and the conventional $\rho=0.5$ SCF lobe. The enlargement of the MI phase also has been studied in single-component bosonic two-leg ladder systems when subjected to staggered  \cite{PhysRevA.98.063612} and uniform \cite{PhysRevA.91.013629, PhysRevA.95.063601} magnetic flux. This phenomenon occurs because the magnetic field has the tendency to localize the single-particle dynamics even for single-component bosonic systems, which makes the system enter an atomic localized phase easier. Remind that since the Hamiltonian is periodic with $\phi$, the 2MI and SCF lobe would grow in size first, and then reduce to satisfy the periodicity. 

Notice that the enlargement is also observed in the strong-coupling expansion calculation revealing that the CGMF can still capture the primary physics in the presence of the magnetic field.  In the meantime, a good agreement can be found between the CGMF and strong-coupling expansion in the 2MI-2SF phase boundary for small values of the hopping amplitude. However, as the hopping amplitude increases, the disparity between the two methods becomes more obvious, mostly due to the limitations of the perturbative theory and the mean-field approximation. 

To our surprise, as depicted in Fig.~\ref{Fig4}(c), another aspect is that three loophole SCF phases can be observed at $J/K=1$ when the magnetic field is sufficiently large. These three loophole SCF phases exhibit the same averaged single-component particle numbers per site as that of the loophole SCF phases in Fig.~\ref{Fig2}(a) where $J/K=0.1$ and $\phi=0$. It can be understood by the fact that the localization of the single-particle dynamics which is induced by a large magnetic flux indicates that the effective $J$ is decreased.  This means that the effective $J/K$ is decreased accordingly. As discussed in section \ref{RA}, if the effective $J/K$ is small enough, the loophole SCF phases can then emerge. 
 
\section{Conclusions}\label{Conclusions}
In summary, we have mainly studied the ground-state phase diagram of the non-hard-core two-component Bose-Hubbard model with and without a uniform magnetic flux on the two-leg ladder by using the cluster Gutzwiller mean-filed method. We have shown that such a system possesses several interesting loophole SCF phases at a sufficient small intra- to inter-leg hopping ratio when there is no magnetic field.  Furthermore, these loophole SCF phases also can persist for a large magnetic flux even if the ratio is not so small. Moreover, the presence of the magnetic flux leads to an enlargement of the Mott insulator lobe and the conventional SCF lobe. We are of the opinion that the model we have examined serves as an example for understanding the fundamental physics of lattice gases coupled to gauge fields. Specifically, it has the potential to inspire experimental investigations into two-leg ladder two-component bosonic systems when subjected to gauge fields.
\section{ACKNOWLEDGEMENTS}
This research is supported by the Scientific Research Fund of Zhejiang Provincial Education Department under Grant No. Y202248878, the Ph.D. research Startup Foundation of Wenzhou University under Grant No. KZ214001P05, the open project of the state key laboratory of surface physics in Fudan University under Grant No. KF2022$\_$06, and Scientific Research Fund for Distinguished Young Scholars of the Education Department of Anhui Province under Grant No. 2022AH020008.

\appendix
\setcounter{figure}{0}

\section{The maximum single-component boson occupy number}\label{A}
\begin{figure}[htbp] 
   \centering
   \includegraphics[width=0.4\textwidth]{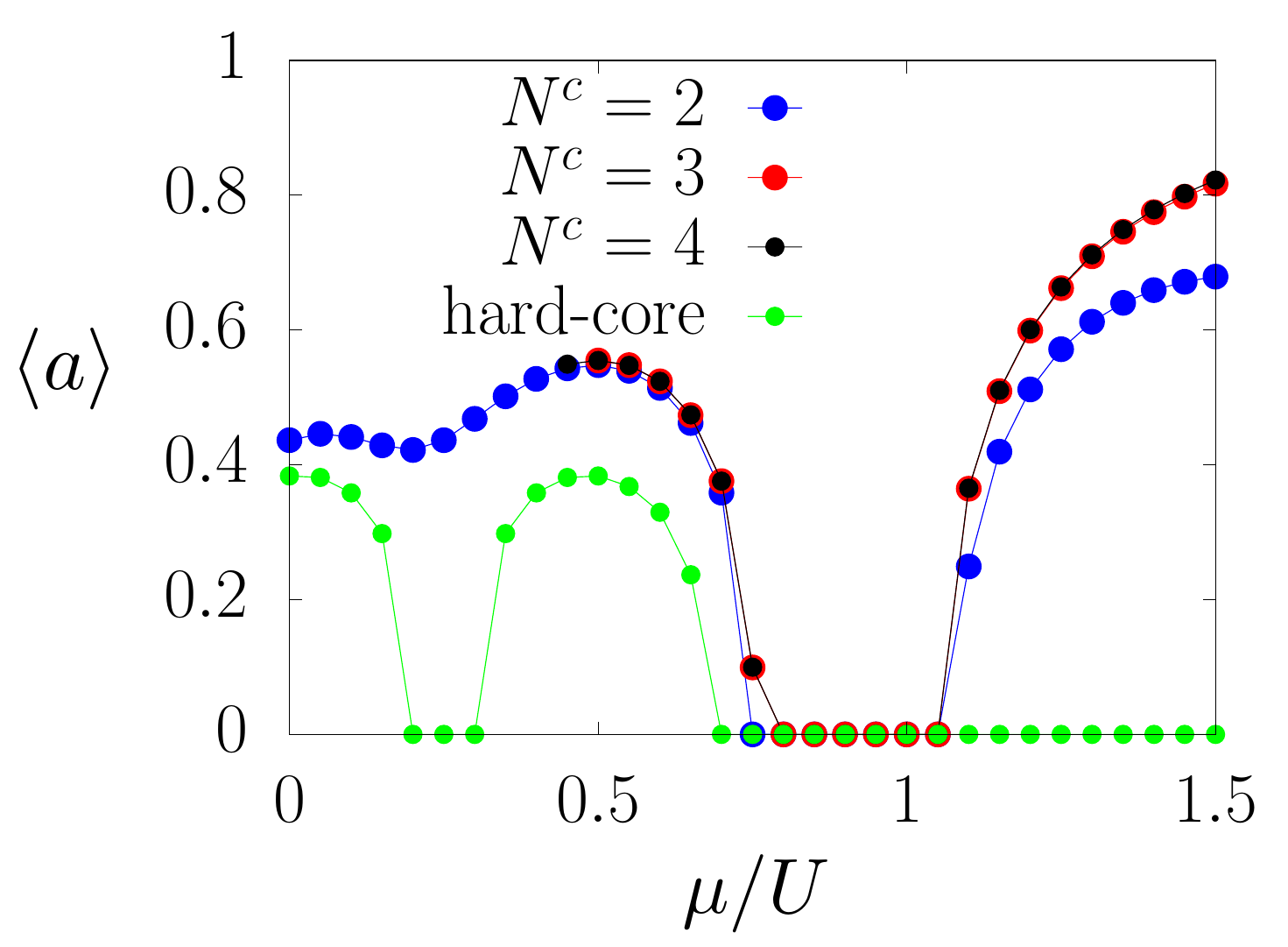} 
   \caption{(Color online) Order parameter $\langle a \rangle$ as a function of $\mu/U$ for different values of $N^c$. The model parameters are set as $J/K=1$, $J/U=0.06$, $U_{ab}/U=0.5$, and $\phi=0$ ($U$ is the energy unit). }
   \label{FigA}
\end{figure}
 The maximum boson occupancy number cutoff refers to the maximum number of bosons that can occupy a single site. Unlike fermions, which are subject to the Pauli exclusion principle and can only have no more than one particle per state, bosons can occupy the same state in unlimited numbers. In numerical methods, a cutoff is often introduced to limit the maximum number of bosons in a single site. This is done to make the calculations tractable. The specific value of the cutoff can vary depending on the system being studied and the computational resources available. In Fig.~\ref{FigA}, we have tested the boson $a$ superfluid order parameter for several single component boson number cutoff $N^c$.  We remind that the two-component bosons $a$ and $b$ share the same value of $N^c$. The relative difference in the value of $\langle a \rangle$ between $N^c=3$ and $4$ in Fig.~\ref{FigA} is less than about $0.7\%$. For this reason, $N^c=3$ is good enough when $\mu/U<1.50$. In the actual calculations, we set $N^c=4$ when $1.0<\mu/U<1.5$ and $N^c=3$ for other values of $\mu/U$. From Fig.~\ref{FigA}, we can also learn that the hard-core and non-hard-core bosons have a huge difference even at a low chemical potential. 
 
 \section{Calculation of the interspecies entanglement entropy}\label{B}
 As discussed in the main text, the interspecies entanglement entropy $S$ can distinguish between an SCF state and a 2MI state. Here, we give some details about how we extract $S$ from the cluster ground state $|C\rangle$. $|C\rangle$ can be expressed in terms of the many-site Fock states of the considered plaquette,
 \begin{equation}
|C\rangle=\sum_k c_{m_1^a,m_1^b,\cdots,m_4^a,m_4^b}^0 |m_1^a,m_1^b,\cdots,m_4^a,m_4^b\rangle,
\end{equation}
where $m_s^a=0,1,2,\cdots N^c$ and $m_s^b=0,1,2,\cdots N^c$ are the number of boson $a$ and $b$ occupied on the plaquette site $s$ respectively. The coefficients $c_{m_1^a,m_1^b,\cdots,m_4^a,m_4^b}^0$ can be treated as an element of a tensor $c^0$ with eight indices. In mathematics, tensor $c^0$ can be reshaped to a matrix $F$ with only two indices. As a example, suppose $f_{i,j}$ is a element of matrix $F$, then 
\begin{equation}
f_{i,j}=c_{m_1^a,m_1^b,\cdots,m_4^a,m_4^b}^0,
\end{equation}
where the indices satisfying,
\begin{equation}
\begin{aligned}
i &=1+m_1^a+m_2^a\cdot(N^c+1) \\ &+m_3^a\cdot(N^c+1)^2+m_4^a\cdot(N^c+1)^3, \\
j &=1+m_1^b+m_2^b\cdot(N^c+1) \\ &+m_3^b\cdot(N^c+1)^2+m_4^b\cdot(N^c+1)^3.
\end{aligned}
\end{equation}
The reduced density matrix $\widetilde{\rho}$, obtained by tracing out the $b$ species, can be written as $\widetilde{\rho} = \text{tr}_b |C\rangle \langle C| = F \cdot F^T$.  $S$ can then be determined by utilizing the eigenvalues of $\widetilde{\rho}$. Numerically, a more efficient approach to evaluate $S$ is directly to perform singular value decomposition on $F$ and obtain its singular values $q_k$. In this case, $S$ can be calculated as $S = -\sum_k |q_k|^2 \ln |q_k|^2$.

\bibliography{Manuscript}

\begin{thebibliography}{10}
\expandafter\ifx\csname url\endcsname\relax
  \def\url#1{\texttt{#1}}\fi
\expandafter\ifx\csname urlprefix\endcsname\relax\def\urlprefix{URL }\fi
\expandafter\ifx\csname href\endcsname\relax
  \def\href#1#2{#2} \def\path#1{#1}\fi

\bibitem{PhysRevA.77.011603}
J.~Catani, L.~De~Sarlo, G.~Barontini, F.~Minardi, M.~Inguscio, Degenerate
  bose-bose mixture in a three-dimensional optical lattice, Phys. Rev. A 77
  (2008) 011603.
\newblock \href {https://doi.org/10.1103/PhysRevA.77.011603}
  {\path{doi:10.1103/PhysRevA.77.011603}}.

\bibitem{PhysRevLett.100.210402}
G.~Thalhammer, G.~Barontini, L.~De~Sarlo, J.~Catani, F.~Minardi, M.~Inguscio,
  Double species bose-einstein condensate with tunable interspecies
  interactions, Phys. Rev. Lett. 100 (2008) 210402.
\newblock \href {https://doi.org/10.1103/PhysRevLett.100.210402}
  {\path{doi:10.1103/PhysRevLett.100.210402}}.

\bibitem{PhysRevLett.105.045303}
B.~Gadway, D.~Pertot, R.~Reimann, D.~Schneble, Superfluidity of interacting
  bosonic mixtures in optical lattices, Phys. Rev. Lett. 105 (2010) 045303.
\newblock \href {https://doi.org/10.1103/PhysRevLett.105.045303}
  {\path{doi:10.1103/PhysRevLett.105.045303}}.

\bibitem{PhysRevA.98.051602}
F.~Sch{\"a}fer, N.~Mizukami, P.~Yu, S.~Koibuchi, A.~Bouscal, Y.~Takahashi,
  Experimental realization of ultracold yb-$^{7}\mathrm{Li}$ mixtures in mixed
  dimensions, Phys. Rev. A 98 (2018) 051602.
\newblock \href {https://doi.org/10.1103/PhysRevA.98.051602}
  {\path{doi:10.1103/PhysRevA.98.051602}}.

\bibitem{science.1122318}
M.~W. Zwierlein, A.~Schirotzek, C.~H. Schunck, W.~Ketterle, Fermionic
  superfluidity with imbalanced spin populations, Science 311~(5760) (2006)
  492--496.
\newblock \href {https://doi.org/10.1126/science.1122318}
  {\path{doi:10.1126/science.1122318}}.

\bibitem{science.1122876}
G.~B. Partridge, W.~Li, R.~I. Kamar, Y.~an~Liao, R.~G. Hulet, Pairing and phase
  separation in a polarized fermi gas, Science 311~(5760) (2006) 503--505.
\newblock \href {https://doi.org/10.1126/science.1122876}
  {\path{doi:10.1126/science.1122876}}.

\bibitem{PhysRevLett.97.030401}
Y.~Shin, M.~W. Zwierlein, C.~H. Schunck, A.~Schirotzek, W.~Ketterle,
  Observation of phase separation in a strongly interacting imbalanced fermi
  gas, Phys. Rev. Lett. 97 (2006) 030401.
\newblock \href {https://doi.org/10.1103/PhysRevLett.97.030401}
  {\path{doi:10.1103/PhysRevLett.97.030401}}.

\bibitem{PhysRevLett.97.190407}
G.~B. Partridge, W.~Li, Y.~A. Liao, R.~G. Hulet, M.~Haque, H.~T.~C. Stoof,
  Deformation of a trapped fermi gas with unequal spin populations, Phys. Rev.
  Lett. 97 (2006) 190407.
\newblock \href {https://doi.org/10.1103/PhysRevLett.97.190407}
  {\path{doi:10.1103/PhysRevLett.97.190407}}.

\bibitem{Nature7179}
Y.-i. Shin, C.~H. Schunck, A.~Schirotzek, W.~Ketterle, Phase diagram of a
  two-component fermi gas with resonant interactions, Nature 451~(7179) (2008)
  689--693.
\newblock \href {https://doi.org/10.1038/nature06473}
  {\path{doi:10.1038/nature06473}}.

\bibitem{science.1077386}
G.~Modugno, G.~Roati, F.~Riboli, F.~Ferlaino, R.~J. Brecha, M.~Inguscio,
  Collapse of a degenerate fermi gas, Science 297~(5590) (2002) 2240--2243.
\newblock \href {https://doi.org/10.1126/science.1077386}
  {\path{doi:10.1126/science.1077386}}.

\bibitem{PhysRevLett.92.140405}
F.~Ferlaino, E.~de~Mirandes, G.~Roati, G.~Modugno, M.~Inguscio, Expansion of a
  fermi gas interacting with a bose-einstein condensate, Phys. Rev. Lett. 92
  (2004) 140405.
\newblock \href {https://doi.org/10.1103/PhysRevLett.92.140405}
  {\path{doi:10.1103/PhysRevLett.92.140405}}.

\bibitem{science.1255380}
I.~Ferrier-Barbut, M.~Delehaye, S.~Laurent, A.~T. Grier, M.~Pierce, B.~S. Rem,
  F.~Chevy, C.~Salomon, A mixture of bose and fermi superfluids, Science
  345~(6200) (2014) 1035--1038.
\newblock \href {https://doi.org/10.1126/science.1255380}
  {\path{doi:10.1126/science.1255380}}.

\bibitem{PhysRevLett.96.180402}
K.~G{\"u}nter, T.~St{\"o}ferle, H.~Moritz, M.~K{\"o}hl, T.~Esslinger,
  Bose-fermi mixtures in a three-dimensional optical lattice, Phys. Rev. Lett.
  96 (2006) 180402.
\newblock \href {https://doi.org/10.1103/PhysRevLett.96.180402}
  {\path{doi:10.1103/PhysRevLett.96.180402}}.

\bibitem{PhysRevLett.102.030408}
T.~Best, S.~Will, U.~Schneider, L.~Hackerm{\"u}ller, D.~van Oosten, I.~Bloch,
  D.-S. L{\"u}hmann, Role of interactions in
  $^{87}\mathrm{Rb}\mathrm{\text{\ensuremath{-}}}^{40}\mathbf{K}$ bose-fermi
  mixtures in a 3d optical lattice, Phys. Rev. Lett. 102 (2009) 030408.
\newblock \href {https://doi.org/10.1103/PhysRevLett.102.030408}
  {\path{doi:10.1103/PhysRevLett.102.030408}}.

\bibitem{PhysRevLett.81.3108}
D.~Jaksch, C.~Bruder, J.~I. Cirac, C.~W. Gardiner, P.~Zoller, Cold bosonic
  atoms in optical lattices, Phys. Rev. Lett. 81 (1998) 3108--3111.
\newblock \href {https://doi.org/10.1103/PhysRevLett.81.3108}
  {\path{doi:10.1103/PhysRevLett.81.3108}}.

\bibitem{PhysRevLett.92.030403}
A.~Kuklov, N.~Prokof'ev, B.~Svistunov, Superfluid-superfluid phase transitions
  in a two-component bose-einstein condensate, Phys. Rev. Lett. 92 (2004)
  030403.
\newblock \href {https://doi.org/10.1103/PhysRevLett.92.030403}
  {\path{doi:10.1103/PhysRevLett.92.030403}}.

\bibitem{PhysRevLett.92.050402}
A.~Kuklov, N.~Prokof'ev, B.~Svistunov, Commensurate two-component bosons in an
  optical lattice: Ground state phase diagram, Phys. Rev. Lett. 92 (2004)
  050402.
\newblock \href {https://doi.org/10.1103/PhysRevLett.92.050402}
  {\path{doi:10.1103/PhysRevLett.92.050402}}.

\bibitem{PhysRevLett.90.100401}
A.~B. Kuklov, B.~V. Svistunov, Counterflow superfluidity of two-species
  ultracold atoms in a commensurate optical lattice, Phys. Rev. Lett. 90 (2003)
  100401.
\newblock \href {https://doi.org/10.1103/PhysRevLett.90.100401}
  {\path{doi:10.1103/PhysRevLett.90.100401}}.

\bibitem{PhysRevA.80.023619}
A.~Hu, L.~Mathey, I.~Danshita, E.~Tiesinga, C.~J. Williams, C.~W. Clark,
  Counterflow and paired superfluidity in one-dimensional bose mixtures in
  optical lattices, Phys. Rev. A 80 (2009) 023619.
\newblock \href {https://doi.org/10.1103/PhysRevA.80.023619}
  {\path{doi:10.1103/PhysRevA.80.023619}}.

\bibitem{SciPost12}
V.~E. Colussi, F.~Caleffi, C.~Menotti, A.~Recati, {Quantum Gutzwiller approach
  for the two-component Bose-Hubbard model}, SciPost Phys. 12 (2022) 111.
\newblock \href {https://doi.org/10.21468/SciPostPhys.12.3.111}
  {\path{doi:10.21468/SciPostPhys.12.3.111}}.

\bibitem{PhysRevLett.125.245301}
Z.~Lin, C.~Liu, Y.~Chen, Novel quantum phases of two-component bosons with pair
  hopping in synthetic dimension, Phys. Rev. Lett. 125 (2020) 245301.
\newblock \href {https://doi.org/10.1103/PhysRevLett.125.245301}
  {\path{doi:10.1103/PhysRevLett.125.245301}}.

\bibitem{PhysRevA.108.013309}
C.~Liu, P.~Chen, L.~He, F.~Xu, Ground-state properties of multicomponent
  bosonic mixtures: A gutzwiller mean-field study, Phys. Rev. A 108 (2023)
  013309.
\newblock \href {https://doi.org/10.1103/PhysRevA.108.013309}
  {\path{doi:10.1103/PhysRevA.108.013309}}.

\bibitem{RevModPhys.83.1405}
M.~A. Cazalilla, R.~Citro, T.~Giamarchi, E.~Orignac, M.~Rigol, One dimensional
  bosons: From condensed matter systems to ultracold gases, Rev. Mod. Phys. 83
  (2011) 1405--1466.
\newblock \href {https://doi.org/10.1103/RevModPhys.83.1405}
  {\path{doi:10.1103/RevModPhys.83.1405}}.

\bibitem{PhysRevA.96.053631}
V.~Penna, A.~Richaud, Two-species boson mixture on a ring: A group-theoretic
  approach to the quantum dynamics of low-energy excitations, Phys. Rev. A 96
  (2017) 053631.
\newblock \href {https://doi.org/10.1103/PhysRevA.96.053631}
  {\path{doi:10.1103/PhysRevA.96.053631}}.

\bibitem{PhysRevB.63.180508}
P.~Donohue, T.~Giamarchi, Mott-superfluid transition in bosonic ladders, Phys.
  Rev. B 63 (2001) 180508.
\newblock \href {https://doi.org/10.1103/PhysRevB.63.180508}
  {\path{doi:10.1103/PhysRevB.63.180508}}.

\bibitem{PhysRevA.96.013620}
A.~Richaud, V.~Penna, Quantum dynamics of bosons in a two-ring ladder:
  Dynamical algebra, vortexlike excitations, and currents, Phys. Rev. A 96
  (2017) 013620.
\newblock \href {https://doi.org/10.1103/PhysRevA.96.013620}
  {\path{doi:10.1103/PhysRevA.96.013620}}.

\bibitem{NaturePhysics}
M.~Atala, M.~Aidelsburger, M.~Lohse, J.~T. Barreiro, B.~Paredes, I.~Bloch,
  Observation of chiral currents with ultracold atoms in bosonic ladders,
  Nature Physics~(8)  588--593.
\newblock \href {https://doi.org/10.1038/nphys2998}
  {\path{doi:10.1038/nphys2998}}.

\bibitem{PhysRevA.92.023618}
H.~Deng, H.~Dai, J.~Huang, X.~Qin, J.~Xu, H.~Zhong, C.~He, C.~Lee, Cluster
  gutzwiller study of the bose-hubbard ladder: Ground-state phase diagram and
  many-body landau-zener dynamics, Phys. Rev. A 92 (2015) 023618.
\newblock \href {https://doi.org/10.1103/PhysRevA.92.023618}
  {\path{doi:10.1103/PhysRevA.92.023618}}.

\bibitem{Tokuno_2014}
A.~Tokuno, A.~Georges, Ground states of a bose–hubbard ladder in an
  artificial magnetic field: field-theoretical approach, New J. Phys. 16~(7)
  (2014) 073005.
\newblock \href {https://doi.org/10.1088/1367-2630/16/7/073005}
  {\path{doi:10.1088/1367-2630/16/7/073005}}.

\bibitem{Strinati_2018}
M.~C. Strinati, F.~Gerbier, L.~Mazza, Spin-gap spectroscopy in a bosonic flux
  ladder, New J. Phys. 20~(1) (2018) 015004.
\newblock \href {https://doi.org/10.1088/1367-2630/aa9ca2}
  {\path{doi:10.1088/1367-2630/aa9ca2}}.

\bibitem{Orignac_2016}
E.~Orignac, R.~Citro, M.~D. Dio, S.~D. Palo, M.-L. Chiofalo, Incommensurate
  phases of a bosonic two-leg ladder under a flux, New J. Phys. 18~(5) (2016)
  055017.
\newblock \href {https://doi.org/10.1088/1367-2630/18/5/055017}
  {\path{doi:10.1088/1367-2630/18/5/055017}}.

\bibitem{PhysRevA.95.063601}
R.~Sachdeva, M.~Singh, T.~Busch, Extended bose-hubbard model for two-leg ladder
  systems in artificial magnetic fields, Phys. Rev. A 95 (2017) 063601.
\newblock \href {https://doi.org/10.1103/PhysRevA.95.063601}
  {\path{doi:10.1103/PhysRevA.95.063601}}.

\bibitem{PhysRevB.91.140406}
M.~Piraud, F.~Heidrich-Meisner, I.~P. McCulloch, S.~Greschner, T.~Vekua,
  U.~Schollw{\"o}ck, Vortex and meissner phases of strongly interacting bosons
  on a two-leg ladder, Phys. Rev. B 91 (2015) 140406.
\newblock \href {https://doi.org/10.1103/PhysRevB.91.140406}
  {\path{doi:10.1103/PhysRevB.91.140406}}.

\bibitem{PhysRevResearch.2.043433}
A.~Haller, A.~S. Matsoukas-Roubeas, Y.~Pan, M.~Rizzi, M.~Burrello, Exploring
  helical phases of matter in bosonic ladders, Phys. Rev. Res. 2 (2020) 043433.
\newblock \href {https://doi.org/10.1103/PhysRevResearch.2.043433}
  {\path{doi:10.1103/PhysRevResearch.2.043433}}.

\bibitem{PhysRevA.98.063612}
R.~Sachdeva, F.~Metz, M.~Singh, T.~Mishra, T.~Busch, Two-leg-ladder
  bose-hubbard models with staggered fluxes, Phys. Rev. A 98 (2018) 063612.
\newblock \href {https://doi.org/10.1103/PhysRevA.98.063612}
  {\path{doi:10.1103/PhysRevA.98.063612}}.

\bibitem{PhysRevA.102.053314}
M.~Buser, C.~Hubig, U.~Schollw\"ock, L.~Tarruell, F.~Heidrich-Meisner,
  Interacting bosonic flux ladders with a synthetic dimension: Ground-state
  phases and quantum quench dynamics, Phys. Rev. A 102 (2020) 053314.
\newblock \href {https://doi.org/10.1103/PhysRevA.102.053314}
  {\path{doi:10.1103/PhysRevA.102.053314}}.

\bibitem{PhysRevA.93.053629}
S.~Uchino, Analytical approach to a bosonic ladder subject to a magnetic field,
  Phys. Rev. A 93 (2016) 053629.
\newblock \href {https://doi.org/10.1103/PhysRevA.93.053629}
  {\path{doi:10.1103/PhysRevA.93.053629}}.

\bibitem{PhysRevA.106.063320}
K.~\ifmmode~\mbox{\c{C}}\else \c{C}\fi{}even, M.~{\"O}. Oktel,
  A.~Kele\ifmmode~\mbox{\c{s}}\else \c{s}\fi{}, Neural-network quantum states
  for a two-leg bose-hubbard ladder under magnetic flux, Phys. Rev. A 106
  (2022) 063320.
\newblock \href {https://doi.org/10.1103/PhysRevA.106.063320}
  {\path{doi:10.1103/PhysRevA.106.063320}}.

\bibitem{PhysRevA.102.043322}
J.-D. Chen, H.-H. Tu, Y.-H. Wu, Z.-F. Xu, Quantum phases of two-component
  bosons on the harper-hofstadter ladder, Phys. Rev. A 102 (2020) 043322.
\newblock \href {https://doi.org/10.1103/PhysRevA.102.043322}
  {\path{doi:10.1103/PhysRevA.102.043322}}.

\bibitem{RevModPhys.82.1225}
C.~Chin, R.~Grimm, P.~Julienne, E.~Tiesinga, Feshbach resonances in ultracold
  gases, Rev. Mod. Phys. 82 (2010) 1225--1286.
\newblock \href {https://doi.org/10.1103/RevModPhys.82.1225}
  {\path{doi:10.1103/RevModPhys.82.1225}}.

\bibitem{PhysRevA.87.043619}
D.-S. L{\"u}hmann, Cluster gutzwiller method for bosonic lattice systems, Phys.
  Rev. A 87 (2013) 043619.
\newblock \href {https://doi.org/10.1103/PhysRevA.87.043619}
  {\path{doi:10.1103/PhysRevA.87.043619}}.

\bibitem{Nat.Phys.7.61}
Y.-A. Chen, S.~D. Huber, S.~Trotzky, I.~Bloch, E.~Altman, Many-body
  landau--zener dynamics in coupled one-dimensional bose liquids, Nature
  Physics~(1)  61--67.
\newblock \href {https://doi.org/10.1038/nphys1801}
  {\path{doi:10.1038/nphys1801}}.

\bibitem{arpack}
R.~B. Lehoucq, D.~C. Sorensen, C.~Yang, ARPACK users' guide: solution of
  large-scale eigenvalue problems with implicitly restarted Arnoldi methods,
  SIAM, Philadelphia, 1998.

\bibitem{PhysRevA.70.061603}
P.~Buonsante, V.~Penna, A.~Vezzani, Fractional-filling loophole insulator
  domains for ultracold bosons in optical superlattices, Phys. Rev. A 70 (2004)
  061603.
\newblock \href {https://doi.org/10.1103/PhysRevA.70.061603}
  {\path{doi:10.1103/PhysRevA.70.061603}}.

\bibitem{PhysRevA.72.031602}
P.~Buonsante, V.~Penna, A.~Vezzani, Fractional-filling mott domains in
  two-dimensional optical superlattices, Phys. Rev. A 72 (2005) 031602.
\newblock \href {https://doi.org/10.1103/PhysRevA.72.031602}
  {\path{doi:10.1103/PhysRevA.72.031602}}.

\bibitem{PhysRevLett.110.260405}
F.~Grusdt, M.~H{\"o}ning, M.~Fleischhauer, Topological edge states in the
  one-dimensional superlattice bose-hubbard model, Phys. Rev. Lett. 110 (2013)
  260405.
\newblock \href {https://doi.org/10.1103/PhysRevLett.110.260405}
  {\path{doi:10.1103/PhysRevLett.110.260405}}.

\bibitem{PhysRevA.72.013614}
P.~Buonsante, A.~Vezzani, Cell strong-coupling perturbative approach to the
  phase diagram of ultracold bosons in optical superlattices, Phys. Rev. A 72
  (2005) 013614.
\newblock \href {https://doi.org/10.1103/PhysRevA.72.013614}
  {\path{doi:10.1103/PhysRevA.72.013614}}.

\bibitem{PhysRevA.94.023634}
L.~Zhang, X.~Qin, Y.~Ke, C.~Lee, Cluster mean-field signature of entanglement
  entropy in bosonic superfluid-insulator transitions, Phys. Rev. A 94 (2016)
  023634.
\newblock \href {https://doi.org/10.1103/PhysRevA.94.023634}
  {\path{doi:10.1103/PhysRevA.94.023634}}.

\bibitem{PhysRevA.91.013629}
A.~Kele\ifmmode~\mbox{\c{s}}\else \c{s}\fi{}, M.~{\"O}. Oktel, Mott transition
  in a two-leg bose-hubbard ladder under an artificial magnetic field, Phys.
  Rev. A 91 (2015) 013629.
\newblock \href {https://doi.org/10.1103/PhysRevA.91.013629}
  {\path{doi:10.1103/PhysRevA.91.013629}}.

\bibitem{PhysRevB.61.12474}
T.~D. K{\"u}hner, S.~R. White, H.~Monien, One-dimensional bose-hubbard model
  with nearest-neighbor interaction, Phys. Rev. B 61 (2000) 12474--12489.
\newblock \href {https://doi.org/10.1103/PhysRevB.61.12474}
  {\path{doi:10.1103/PhysRevB.61.12474}}.

\bibitem{PhysRevB.58.R14741}
T.~D. K{\"u}hner, H.~Monien, Phases of the one-dimensional bose-hubbard model,
  Phys. Rev. B 58 (1998) R14741--R14744.
\newblock \href {https://doi.org/10.1103/PhysRevB.58.R14741}
  {\path{doi:10.1103/PhysRevB.58.R14741}}.

\bibitem{PhysRevA.82.033630}
M.~Iskin, Strong-coupling expansion for the two-species bose-hubbard model,
  Phys. Rev. A 82 (2010) 033630.
\newblock \href {https://doi.org/10.1103/PhysRevA.82.033630}
  {\path{doi:10.1103/PhysRevA.82.033630}}.

\end{thebibliography}
\end{document}